\documentclass[twocolumn]{aastex63}
\usepackage[utf8]{inputenc}
\usepackage{graphicx}
\usepackage[flushleft]{threeparttable}
\usepackage{tabularx}
\usepackage[normalem]{ulem}
\usepackage{amsfonts,amsmath,amssymb}
\usepackage{url}

\shorttitle{$c$-M relation at high mass end}
\shortauthors{Xu et al.}

\begin{document}
\title{Halo Mass-Concentration Relation at High-Mass End}

\author[0000-0002-9587-6683]{Weiwei Xu}
\thanks{Corresponding author: wwxu@pku.edu.cn}
\affiliation{Kavli Institute for Astronomy and Astrophysics, Peking University, Beijing 100871, China}
\author{Huanyuan Shan}
\thanks{Corresponding author: hyshan@shao.ac.cn}
\affiliation{Shanghai Astronomical Observatory, Chinese Academy of Sciences, Shanghai 200030, China}
\affiliation{University of Chinese Academy of Sciences, Beijing 100049, China}
\author{Ran Li}
\affiliation{National Astronomical Observatory, Chinese Academy of Sciences, Beijing 100101, China}
\author{Chunxiang Wang}
\affiliation{National Astronomical Observatory, Chinese Academy of Sciences, Beijing 100101, China}
\author{Linhua Jiang}
\affiliation{Kavli Institute for Astronomy and Astrophysics, Peking University, Beijing 100871, China}
\author{Eric Jullo}
\affiliation{Aix Marseille Univ, CNRS, CNES, LAM, 13388 Marseille, France}
\author{Ginevra Favole}
\affiliation{Institute of Physics, Laboratory of Astrophysics, Ecole Polytechnique F$\acute{e}$d$\acute{e}$rale de Lausanne (EPFL), Observatoire de Sauverny, 1290 Versoix, Switzerland}
\author{Jean-Paul Kneib}
\affiliation{Institute of Physics, Laboratory of Astrophysics, Ecole Polytechnique F$\acute{e}$d$\acute{e}$rale de Lausanne (EPFL), Observatoire de Sauverny, 1290 Versoix, Switzerland}
\author{Chaoli Zhang}
\affiliation{College of Computer Science and Artificial Intelligence, Wenzhou University, 325035 Wenzhou, China}

\begin{abstract}
The concentration-mass ($c$-M) relation encodes the key information of the assembly history of the dark matter halos, however its behavior at the high mass end has not been measured precisely in
observations yet. In this paper, we report the measurement of halo $c$-M relation with galaxy-galaxy lensing method, using shear catalog of the Dark Energy Camera Legacy Survey (DECaLS) Data Release 8, which covers a sky area of $9\,500$ deg$^2$. The
foreground lenses are selected from redMaPPer, LOWZ, and CMASS catalogs, with halo mass range from $10^{13}$ to $10^{15}~\rm{M}_{\odot}$ and redshift range from $z=0.08$ to $z=0.65$. We find that the concentration decreases with the halo mass from $10^{13}$
to $10^{14}~{\rm M}_{\odot}$, but shows a trend of upturn after the pivot point of $\sim10^{14}~{\rm M}_{\odot}$. We fit the measured $c$-M relation with the concentration model
$c(M)=C_0~(\frac{\rm M}{10^{12}~{\rm M}_{\odot}/h})^{-\gamma}~[1+(\frac{\rm{M}}{M_0})^{0.4}]$, 
and get the values ($C_0$, $\gamma$, log$_{10}$($M_0$)) = ($5.119_{-0.185}^{0.183}$, $0.205_{-0.010}^{0.010}$, $14.083_{-0.133}^{0.130}$), and ($4.875_{-0.208}^{0.209}$, $0.221_{-0.010}^{0.010}$,$13.750_{-0.141}^{0.142}$) for halos with $0.08\le z<0.35$ and $0.35\le z<0.65$, respectively. We also show that the model including an upturn is favored over a simple power-law model. Our measurement provides important information for the recent argument of massive cluster formation process.
\end{abstract}

\section{Introduction}

As fundamental building blocks of the dark matter universe, halos are believed to follow a self-similar structure distribution. The most widely used density profile of dark matter halos are proposed by \citet{Navarro1995,Navarro1996,Navarro1997} (hereafter NFW), where the slope of density $\gamma$ equals to 1 at inner part and grows to 3 at outer part. In such a model, one can define the concentration of a halo to be the ratio between the virial radius ($r_{\rm vir}$) and the characteristic radius ($r_{\rm s}$), where the density slope $\gamma(r_{\rm s})=2$. Given the mass and concentration, the density distribution of a halo is determined. 

The halo concentration is confirmed to vary with its redshift and mass both in observations and N-body simulations (e.g., \citealt{McClintock2019,Shan2012,Shan2017,Johnston2007,Cui2018}). 
N-body simulations show that the relation between concentration and mass (called $c$-M relation hereafter) can be used to probe the formation and evolution history of the halo (e.g., \citealt{Zhao2003a, Zhao2003b,Bullock2001}), and the dependence of concentration on the halo mass and redshift can be described with a power-law function, $c=\alpha~(M/M_{\rm pivot})^\beta ~(1+z)^\gamma$ (e.g., \citealt{Duffy2008,Bullock2001,Neto2007,Eke2001,Cui2018}).
Recently, however, it is argued that the power-law may not extend to the high mass end. Some N-body simulations \citep{Klypin2011,Klypin2016,Prada2012,Ishiyama2020} predict the flattening and upturn of the $c$-M relation for massive halos, and the location of the upturn varies with redshift, which may due to the different formation procedure of massive halos at higher redshift.


On observational side, gravitational lensing is the only way to detect the profile 
and mass of the halo directly, without any assumption 
about its hydrodynamic state, symmetry, or profile. 
This makes it a unique way to get tight constraint of the structure and evolution of halos
\citep{Abbott2018, abbott2020,Kwan2017}.
Several previous works have been undertaken to constrain the $c$-M relation with a variety of
data sets, such as the Dark Energy Survey (DES) Science Verfication data \citep{Melchior2017}, DES Year 1 data \citep{McClintock2019}, DES Year 3 \citep{Varga2021}, the Cluster Lensing and Supernova Survey with Hubble data (CLASH, \citealt{Merten2015,Sereno2015}).
Some more works aim to measure the X-ray concentration \citep{Pointecouteau2005, Vikhlinin2006, Gastaldello2007, Sato2000, Buote2007, Comerford2007}. However, they usually describe the $c$-M relation with 
the power-law function, where the concentration decreases with the halo mass in a wide mass range.

In this paper, we perform the measurement of halo $c$-M relation with galaxy-galaxy lensing method, using shear catalog of the Dark Energy Camera Legacy Survey (DECaLS) Data Release 8, which covers a sky area of 9\,500 deg$^2$. The foreground lenses are selected from redMaPPer, LOWZ, and CMASS catalogs, with the range of halo mass $10^{13}-10^{15}~{\rm M}_{\odot}$ and redshift range $0.08-0.65$. Throughout the paper, we take the M$_{200\rm m}$ and $c_{200\rm m}$ as the mass and concentration of the halo, which means the mean density in the halo is 200 times of the matter background density at the same redshift. We make use of different $c$-M models to fit the data and try to investigate whether an upturn exist.

The set of cosmological parameters used is obtained from \citet{Planck2018}, 
the Hubble constant $H_0=67.4$~km~s$^{-1}$~Mpc$^{-1}$, 
the baryon density parameter $\Omega_{\rm b}h^2=0.0224$, 
cold dark matter density parameter $\Omega_{\rm cdm}h^2=0.120$, 
matter fluctuation amplitude $\sigma_8=0.811$, the power index of primordial power spectrum n$_{\rm s}=0.965$,
and matter density parameter $\Omega_{\rm m} = 0.315$.
The structure of this paper is listed as follows. In Sec.~\ref{sec:data}, the source
and lens catalogs are described. In Sec.~\ref{sec:method}, the lensing signal, lensing model, and systematics 
are shown. In Sec.~\ref{sec:result}, we discuss the $c$-M relation measurement and fitting. In Sec.~\ref{sec:discussion} and Sec.~\ref{sec:conclusion}, the discussion and conclusion
are shown, respectively.

\section{Data}
\label{sec:data}
\subsection{Source catalog}
\label{subsec:source_catalog}

The source galaxies used in our measurement are obtained from the Data Release 8 (DR8) of DECaLS survey, which is part of the Dark
Energy Spectroscopic Instrument (DESI) Legacy Imaging
Survey (\citealt{Blum2016,Dey2019}). The sky coverage of DECaLS DR8 is $\sim 9\,500~$deg$^2$ in $grz$ bands. 

In the DECaLS DR8 catalog, the sources from the {\it Tractor} catalog \citep{Lang2014} are divided into five 
kinds of morphologies, including point sources, round exponential galaxies with a 
variable radius, DeVaucouleurs, exponential, and the composite model. The sources 
above $6\sigma$ detection limit in any stack 
are kept as candidates. The ellipticity of galaxy is estimated by a joint fitting on 
the optical images in three bands ($g$, $r$, and $z$). 
Then, we model the multiplicative and additive biases by cross-matching the 
DECaLS sources with external
shear measurements \citep{Phriksee2020, Yao2020, Zu2021}, including Canada-France-Hawaii 
Telescope Stripe 82 \citep{Moraes2014}, Dark Energy Survey
\citep{DES2016}, and Kilo-Degree Survey \citep{Hildebrandt2017} sources. 
 
We use the photometric redshift derived by \citet{Zou2019} with the 
$k$-nearest-neighbour (kNN) algorithm. 
The redshift of a target galaxy is derived with its $k$-nearest-neighbour
in Spectral Energy Distribution (SED) space whose spectroscopic redshift is known.
The photometric redshift (photo-$z$) is obtained from $5$ photometric bands: three optical bands 
($g$, $r$, and $z$), and two infrared bands ($W1$ and $W2$) from 
the Wide-Field Infrared Survey Explorer. We only take samples with $r<23$~mag, resulting in a spectroscopic sample of about $2.2$ million galaxies. 
The characteristics of the final catalog include redshift bias of 
$\Delta z_{\rm norm}=2.4\times10^{-4}$, accuracy of $\sigma_{\Delta z_{\rm norm}}=0.017$, and 
outlier rate of about $5.1\%$. More details are discussed in \citet{Zou2019}.

\subsection{Lens catalogs}

\begin{figure}[t]
    \centering
    \includegraphics[scale=0.45]{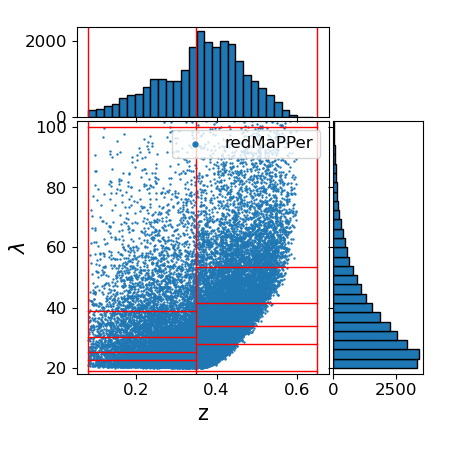}
    \includegraphics[scale=0.45]{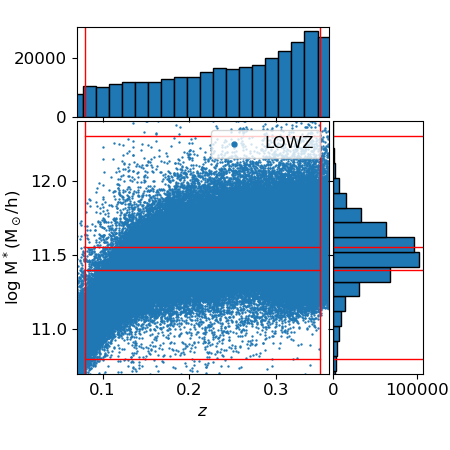}
    \includegraphics[scale=0.45]{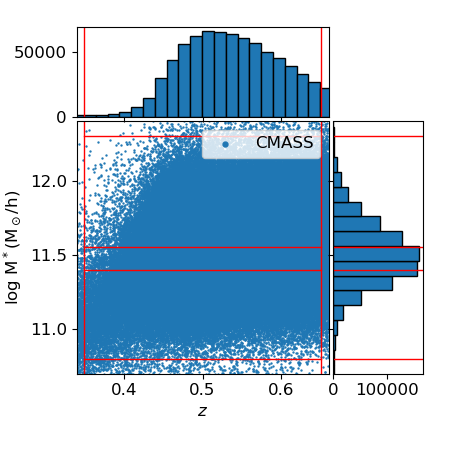}
    \caption{The distribution of redMaPPer, LOWZ, and CMASS samples in 
    the redshift and mass proxy parameter space. The mass proxy is the cluster richness 
    for redMaPPer halos and the logarithm of the stellar mass for LOWZ and CMASS halos. 
    The red solid lines label out thresholds of bins.}
    \label{fig:bins}
\end{figure}

We use the redMaPPer cluster catalog v6.3\footnote{http://risa.stanford.edu/redmapper/} as the lens catalog, which is obtained from the Sloan Digital Sky Survey (SDSS) DR8 \citep{Rykoff2014}.
The catalog includes $26\,111$ clusters with richness $\lambda>19$. 
In order to get the evolution of $c$-M relation with redshift and mass, the redMaPPer cluster sample is separated into multiple 
redshift and richness bins, as shown in Fig.~\ref{fig:bins} and Tab.~\ref{tab:bins}. 
The outliers with $\lambda>100$ are removed to get the main characters of the whole sample without 
bias from a few outliers. In this step, $375$ most massive clusters are removed, which account for $1.4\%$ of the full sample.
Then, the sample is separated into two redshift parts, i.e., low-$z$ sample ($0.08\le z< 0.35$, hereafter $z1$ sample) and 
high-$z$ sample ($0.35\le z<0.65$, hereafter $z2$ sample).
The number ratio of clusters in these two redshift samples are $40.3\%$ and $59.7\%$, respectively. There are relatively more clusters in the high-$z$ sample, this can partly compensate the weakness of remote cluster signal.
The cluster richness are used as the mass proxy to separate the massive and less massive clusters. 
Each redshift sample is further divided into $5$ richness bins (named $\lambda4-\lambda8$ bins), 
with similar cluster number in each bin.
This similarity of cluster number can control unnecessary bias when we measure and compare the weak 
lensing signal of each bin. Using the galaxy-galaxy lensing measurement, 
we can further obtain the density profile of clusters with different halo mass at different redshifts.

Besides redMaPPer catalog, we also utilize the LOWZ and CMASS catalogs from SDSS-III BOSS DR10 \citep{Ahn2014} to 
constrain the $c$-M relation at the low mass region. 
The redshift of LOWZ and CMASS halos are constraint within $0.08-0.35$ and $0.35-0.65$, respectively, 
to match with the redshift bins of redMaPPer halos. For these two catalogs,
we take the stellar mass as the halo mass
proxy and separate each catalog into $3$ mass bins (named $\lambda 1-\lambda 3$ bins),
as shown in Fig.~\ref{fig:bins} and Tab.~\ref{tab:bins}. 
To minimize the effect of outliers, we remove halos with stellar mass or redshift out of the listed range. 

\begin{table}[t]
    \centering
    \footnotesize
    \begin{tabular}{c|c|c|c|c|c}
    \hline  
    Catalog & bin &$z$ & log$_{10}$(M*) or $\lambda$ & N. & P.($\%$)  \\
    \hline
    LOWZ   & $\lambda1z1$ & $0.08$ - $0.35$ & $10.80$ - $11.40$ &  $97\,280$ & $33.4$ \\
           & $\lambda2z1$ & $0.08$ - $0.35$ & $11.40$ - $11.55$ & $105\,186$ & $36.1$ \\
           & $\lambda3z1$ & $0.08$ - $0.35$ & $11.55$ - $12.30$ &  $88\,648$ & $30.5$ \\
    \hline
    CMASS  & $\lambda1z2$ & $0.35$ - $0.65$ & $10.80$ - $11.40$ & $230\,280$ & $31.5$ \\
           & $\lambda2z2$ & $0.35$ - $0.65$ & $11.40$ - $11.55$ & $225\,026$ & $30.7$ \\
           & $\lambda3z2$ & $0.35$ - $0.65$ & $11.55$ - $12.30$ & $276\,716$ & $37.8$ \\
    \hline
    redMP. & $\lambda4z1$ & $0.08$ - $0.35$ & $19.00$ - $22.40$ &   $2\,078$ & $8.0$\\
           & $\lambda5z1$ & $0.08$ - $0.35$ & $22.40$ - $25.31$ &   $2\,067$ & $7.9$\\
           & $\lambda6z1$ & $0.08$ - $0.35$ & $25.31$ - $30.15$ &   $2\,087$ & $8.0$\\
           & $\lambda7z1$ & $0.08$ - $0.35$ & $30.15$ - $38.83$ &   $2\,074$ & $7.9$\\
           & $\lambda8z1$ & $0.08$ - $0.35$ & $38.83$ - $100~~$  &   $2\,076$ & $8.0$\\
    \hline
    redMP. & $\lambda4z2$ & $0.35$ - $0.65$ & $19.00$ - $27.72$ &   $3\,058$ & $11.7$ \\
           & $\lambda5z2$ & $0.35$ - $0.65$ & $27.72$ - $33.82$ &   $3\,078$ & $11.8$ \\
           & $\lambda6z2$ & $0.35$ - $0.65$ & $33.82$ - $41.51$ &   $3\,067$ & $11.7$ \\
           & $\lambda7z2$ & $0.35$ - $0.65$ & $41.51$ - $53.37$ &   $3\,081$ & $11.8$ \\
           & $\lambda8z2$ & $0.35$ - $0.65$ & $53.37$ - $100~~$ &   $3\,070$ & $11.8$ \\
    \hline  
    \end{tabular}
    \caption{The details of the redshift and mass proxy thresholds for LOWZ, CMASS,
    and redMaPPer (shown as 'redMP.' in the table for short) subsamples. 
    The first column shows the catalog name. The 
    second column lists the given name of each bin. The thresholds of redshift and mass proxy 
    are shown in the 3rd and 4th columns.
    The mass proxy is the logarithm value of stellar mass for LOWZ and CMASS catalogs, and the richness for redMaPPer.
    In last two columns, the included number in this bin and its percentage of the whole catalog are listed.}
    \label{tab:bins}
\end{table}

\section{Method}
\label{sec:method}

\subsection{The lensing signal}

The gravitational well of the foreground halo produces a tangential shear of the 
source around the foreground halo, which stretches 
and aligns the source images along the tangential direction. Thus, the projected mass density of lens, $\Sigma$, is related to the azimuthally averaged tangential shear at projected radius $R$. 
The relation is
\begin{equation}
\gamma_{\rm T} = \frac{\bar\Sigma(<R)-\bar\Sigma(R)}{\Sigma_{\rm crit}} \equiv \frac{\Delta\Sigma(R)}{\Sigma_{\rm crit}}.
\end{equation}

The $\Delta\Sigma$(R) is widely used to show the 
weak lensing signal, the reduced shear ($g$), which is defined as 
$g=\gamma/(1-\kappa)$, where $\kappa$ is the dimensionless surface mass 
density and defined as $\kappa={\Sigma(R)}/
{\Sigma_{\rm crit}}$.
In this paper, we also use $\Delta\Sigma$(R) to show the detected signal. 

The critical surface mass density, $\Sigma_{\rm crit}$, is defined as
\begin{equation}
\Sigma_{\rm crit} (z_{\rm l}, z_{\rm s})= \frac{c^2}{4\pi G}~ \frac{D_{\rm s}}{{D_{\rm l}} D_{\rm ls}},
\label{equ:sigma_crit}
\end{equation}
where the $D_{\rm l}$, $D_{\rm s}$, and $D_{\rm ls}$ are the angular diameter 
distance to the lens, to source, and between the lens and source, respectively, and the $c$ here is the constant of light velocity in vacuum. 
The $\Sigma_{\rm crit}$ shows how the geometry of the lens-source system
modulates the induced shear signal. 

To obtain $\Delta\Sigma$, we stack lens-source pairs in $10$ logarithmic co-moving radius 
of $0.3-5$~Mpc in this work. Only sources with 
$z_{\rm s}>z_{\rm l}+0.1$ are considered, to avoid the mis-classification caused by the redshift uncertainty. With this set of parameters, as listed in Tab.~\ref{tab:swot_para}, $\Delta\Sigma(R)$ is estimated for a given set of lenses using
\begin{equation}
\Delta\Sigma(R)=\frac{\sum_{\rm ls}w_{\rm ls}~\gamma_t^{\rm ls}~\Sigma_{\rm crit}}{\sum_{\rm ls}w_{\rm ls}}\,,
\end{equation}
where $\gamma_t^{\rm ls}$ is the tangential shear, $w_{\rm ls}=w_{\rm n}\Sigma_{\rm crit}^{-2}$, and $w_{\rm n}$ is the weight factor introduced to account for intrinsic scatter in ellipticity and the error of shape measurement \citep{Miller2007,Miller2013}. The $w_{\rm n}$ used in this work is defined as $w_{\rm n}=1/(\sigma^2_{\epsilon}+\sigma^2_{\rm e})$. The $\sigma_{\epsilon}$ is the intrinsic ellipticity dispersion derived from the whole galaxy catalogue, and taken as $0.27$ \citep{Giblin2021}. $\sigma_{\rm e}$ is the error of the ellipticity measurement defined in \citet{Hoekstra2002}.

The lensing signal is recalibrated as
\begin{equation}
\Delta\Sigma^{\rm cal}(R)=\frac{\Delta\Sigma(R)}{1+K(z_{\rm l})},
\end{equation}
and \begin{equation}
1+K(z_{\rm l})=\frac{\sum_{\rm ls}w_{\rm ls}~(1+m)}{\sum_{\rm ls}w_{\rm ls}},
\end{equation}
where $m$ is the multiplicative error as described in Sec.~\ref{subsec:source_catalog}.

We use the software $SWOT$\footnote{http://jeancoupon.com/swot} \citep{Coupon2012} to detect the stacked 
signal. It is a fast tree-code to compute the two-point correlations, histograms, and galaxy-galaxy 
lensing signal from large datasets.
The $theta$ projection is taken to measure the signal. The software can be parallelized for a maximum computational efficiency.
We estimate the statistical error with a Jackknife resampling of $64$ sub-regions with equal area, and remove one sub-sample at a time for each Jackknife realisation. Refer to Tab.~\ref{tab:swot_para} for $SWOT$ parameter setting.

\begin{table}[t]
    \centering
    \begin{tabular}{c|c|l}
    \hline
    Para. & value & Meaning \\
    \hline
        corr                & gglens    & Type of correlation\\
        range               & 0.1,~7     & Correlation range (in the unit of Mpc/h)\\
        nbins               & 15        & Number of bins \\
        err                 & Jackknife & Resampling method\\
        nsub                & 64	    & Number of resampling subvolumes \\
        H$_0$                  & 67.4      & Hubble parameter\\
        $\Omega_{\rm m}$    & 0.315     & Relative matter density\\
        $\Omega_{\rm L}$    & 0.684     & Relative energy density\\
        $\Delta$            & 0.1       & Minimum redshift difference \\
                            &           & between the source and the lens\\
        proj                & como      & Projection\\
    \hline
    \end{tabular}
    \caption{Parameter setting of $SWOT$. The name, value and physical meaning of parameters are listed in three columns in sequence.}
    \label{tab:swot_para}
\end{table}

\subsection{The lensing model}

We fit the observation with a comprehensive model, including
the central halo, 
the miscentered halo,
and nearby halo term with offset.

Firstly, we use the NFW profile to estimate the contribution from the central halo.
In addition, we take into account the miscentering effect, which comes from the inaccurate determination of 
halo center, and can reduce the central signal greatly \citep{Johnston2007}.
For a cluster miscentered by the distance $R_{\rm mis}$, 
the surface mass density is $\Sigma_{\rm mis}=\int^{2\pi}_0 {{\rm d}\theta}~\Sigma(\sqrt{R^2+R_{\rm mis}^2+2RR_{\rm mis} \rm{cos} \theta})/{2\pi}$. We assume a Gamma
profile for the miscentering of the stacked signal \citep{McClintock2019}. The miscentering effect is characterized by two parameters, $f_{\rm mis}$ and $r_{\rm mis}$, representing the fraction of offset halos and the offset distance, respectively. 
Finally, we use the two halo term to indicate the signal from nearby halos, 
which dominates at the cluster outskirt.
The contribution from the two halo term is estimated from the non-linear 
scaling of the matter power spectrum as a function of redshift with the $Halofit$ model, using the $CAMB$ package\footnote{https://github.com/cmbant/CAMB}.
More details about the model refer to the $Cluster\_toolkit$ package\footnote{https://cluster-toolkit.readthedocs.io/en/latest/} 
\citep{Smith2003,Eisenstein1998,Takahashi2012}. 
Thus, the whole model is, 
\begin{equation}
\begin{aligned}
\Delta\Sigma(R) = 
 & ~(1-f_{\rm mis})~\Delta\Sigma_{\rm NFW}(R) 
 + f_{\rm mis}~\Delta\Sigma_{\rm mis}(R) \\
 & +\Delta\Sigma_{\rm 2h}(R).
 \end{aligned}
\end{equation}

\subsection{Systematics}

In the model fitting, there are multiplicative corrections \citep{McClintock2019} necessary to consider,
including the boost factor ($\mathcal{B}(R)$), reduced shear ($\mathcal{G}(R)$), and photo-$z$ bias ($\delta$). With these corrections, the observed signal is $\Delta\Sigma^{\rm cal}(R) = (1+\delta) ~\mathcal{G}(R)/\mathcal{B}(R)~\Delta\Sigma_{\rm model}$. 

\subsubsection{Boost factor}

The boost effect \citep{Sheldon2004,Mandelbaum2006} 
comes from the membership dilution biases when some foreground or member galaxies are mis-classified as background sources. If the fraction of mis-classified member galaxies for 
the cluster is $f_{\rm cl}$, the boost factor, $\mathcal{B}(R)=(1-f_{\rm cl})^{-1}$, is used to 
correct the diluted signal. In this work, we use the boost factor model (Eq.~\ref{eq:boost_model}) referring to \citet{McClintock2019},
which is constructed from the NFW profile and characterised with $B_0$ and scale radius $R_s$. We use the typical values of $B_0 = 0.1$ and $R_{\rm s} = 1.0$~Mpc/$h$.

\begin{equation}
  \label{eq:boost_model}
    \mathcal{B}_{\rm model}(R) = 1+B_0 \frac{1-F(x)}{x^2-1},~~     (x=R/R_s)
\end{equation}
\begin{align}
  \label{eq:boost_model2}
  F(x) = \left\{
  \begin{array}{lr}
    \frac{\tan^{-1}\sqrt{x^2-1}}{\sqrt{x^2-1}} & ,~~ (x > 1)\\
    1 & , ~~(x = 1)\\
    \frac{\tanh^{-1}\sqrt{1-x^2}}{\sqrt{1-x^2}} & ,~~(x < 1)
  \end{array}
  \right.\,.
\end{align}

\subsubsection{Reduced shear error}

The reduced shear error comes from the fact that the measured signal is the reduced 
shear ($g$), instead of the shear ($\gamma$). This can be corrected by multiplying the model with
\begin{equation}
\mathcal{G}(R) = \frac{1}{1-\kappa} =\frac{1}{1-\Sigma(R)~\Sigma_{\rm crit}^{-1}}.
\end{equation}
Here, the $\Sigma(R)$ includes the contribution of the central halo, nearby halos, and miscentering effect.

\subsubsection{Photo-$z$ bias}

Besides the corrections mentioned above, we also consider the
systematic uncertainties of photo-$z$ ($\delta$).

The redshift of source is the photometric redshift estimated by the kNN algorithm (see Sec.~\ref{subsec:source_catalog}), 
and local linear regression is used as described in \citet{Zou2019}. 
For each galaxy, our kNN photo-$z$ algorithm provides a Gaussian 
estimation of the photo-$z$ error. We further apply this Gaussian scatter for each galaxy to obtain the 
probability distribution function (PDF) of redshift, i.e., $p_{\rm photo}(z_{\rm s})$.
For each galaxy, we estimate the effective critical surface density with the redshift PDF as
\begin{equation}
\langle\Sigma^{-1}_{\rm crit}\rangle_{i,j}=\int {\rm d} z_{\rm s}~ p_{\rm phot}(z_{{\rm s},i}) \Sigma^{-1}_{\rm crit}(z_{{\rm l},j}, z_{{\rm s}, i}),
\end{equation}
where $i$ and $j$ indicate the source and lens in a lens-source pair.
For the case with $z_{\rm s}<z_{\rm l}$ or the offset $>0.3$~Mpc, the value of $\langle \Sigma^{-1}_{\rm crit} \rangle_{i,j}=0$. 
We only consider the signal from the lens-source pair when their projected distance is smaller 
than $0.3$~Mpc and when the background source has a higher 
redshift than the foreground lens. 
Furthermore, we compare the kNN photo-$z$ catalog of the photometric redshift 
(PDF: $p^{\rm kNN}_{\rm photo}(z_{\rm s})$) with a large reliable photo-$z$
catalog (PDF: $p^{\rm true}_{\rm photo}(z_{\rm s})$) obtained from UDS HSC+SPLASH 
\citep{Mehta2018}, ECDF-S \citep{Cardamone2010}, 
CFHTLS Deep + WIRDS \citep{Bielby2010}, and COSMOS \citep{Laigle2016}. 
As a result, the bias of photo-$z$ is given as
\begin{equation}
1+\delta=\frac{\Sigma^{-1}_{\rm crit, kNN}}{\Sigma^{-1}_{\rm crit, true}}
\end{equation}

\section{Result}
\label{sec:result}

The final parameters are estimated by the Markov Chain Monte Carlo (MCMC) fitting method 
using the $EMCEE$ package\footnote{https://emcee. readthedocs.io/en/stable} \citep{emcee}. 
We use $50$ chains with the original length of 
$30\,000$ steps each, and discard the first $10\,000$ steps, similar with the implementation in \citet{Yang2020}. The first part of chains are discarded to avoid the effect from first guess of parameters.
The relation between the mass and concentration, obtained from the stacked signal of halos in 
redMaPPer, CMASS, and LOWZ catalogs, are shown in Fig.~\ref{fig:cmRelation}. 
All these three lens samples are divided into two redshift bins. And in each
redshift bin, there are $5$ mass bins for redMaPPer clusters, and $3$ mass
bins for CMASS, LOWZ galaxies, respectively. The criteria are shown in Tab.~\ref{tab:bins}. 

\subsection{$c$-M relation measurement}
\label{sec:fit_steps}

For the performance of signal, only signal with the error smaller than $100\%$ are used. We measure the signal in $15$ logarithmic bins in the radius of $0.1-7$~Mpc, and only fit the $10$ bins in $0.3-5$~Mpc, to avoid the contribution of central galaxy and nearby halos. We use maximum likelihood fitting method to calculate best parameters in the fitting. There are
$4$ parameters to estimate, i.e., halo mass (M$_{200\rm m}$), 
concentration ($c_{200\rm m}$), offset distance ($r_{\rm mis}$), ratio of halo with offset ($f_{\rm mis}$). 

First, we fit the model with MCMC method. Gaussian priors are assumed for halo mass and concentration, whose central value and full width at half maximum (FWHM)
are listed in Tab.~\ref{tab:prior}. The $r_{\rm mis}$ and $f_{\rm mis}$ are assumed flat priors
within $0-1$~Mpc and $0-1$, respectively, with initial starting value both set as $0.2$. Second, the four parameters are fitted with MCMC method again,
and all the four priors are assumed to be a Gaussian distribution, with the
central value and the FWHM as the best fitted results in the previous step. 
Third, the $r_{\rm mis}$ and $f_{\rm mis}$ are fixed to the best fitted value obtained 
from the second step. And the mass and concentration of halo are fitted with MCMC method,
with priors listed in Tab.~\ref{tab:prior}. Fourth, the mass and concentration are
assigned gaussian priors, with the central value and FWHM estimated from the third step.
Finally, the halo mass is fixed to the value obtained in the 4th step, and the concentration
is fitted with MCMC method, with the gaussian prior from the 4th step.

\begin{table}[t]
    \centering
    \begin{tabular}{c|c|c}
    \hline
    Samp.       & log$_{10}$(M)     & $c$        \\
    \hline
    redMP.       & [$14.5, 1.0$] & [$4.0, 0.5$] \\
    LOWZ, CMASS  & [$13.5, 1.0$] & [$4.5, 0.5$] \\
    \hline
    \end{tabular}
    \caption{Gaussian priors of the mass and concentration of halos from redMaPPer, LOWZ, and CMASS catalogs. For each sample, we list the central value and FWHM of the gaussian function.}
    \label{tab:prior}
\end{table}

\begin{figure*}[t]
    \centering     
    Step 1  ~~~~~~~~~~~~~~~~~~~~~~~~~~~~~~~~~~~~~~~~~~~~~~~~~~~~~~~~~~~ Step 2\\
    \includegraphics[scale=0.45]{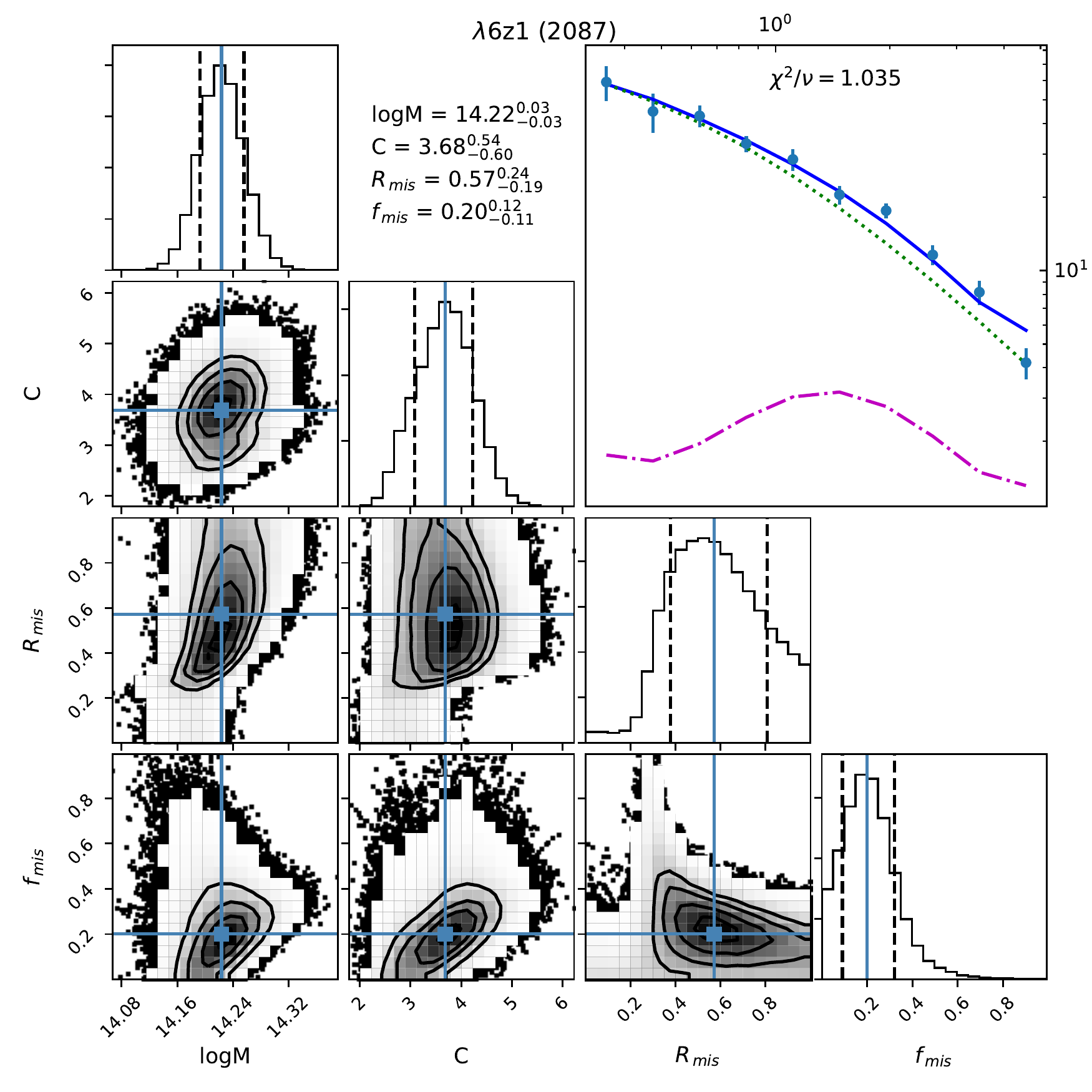}    
    \includegraphics[scale=0.45]{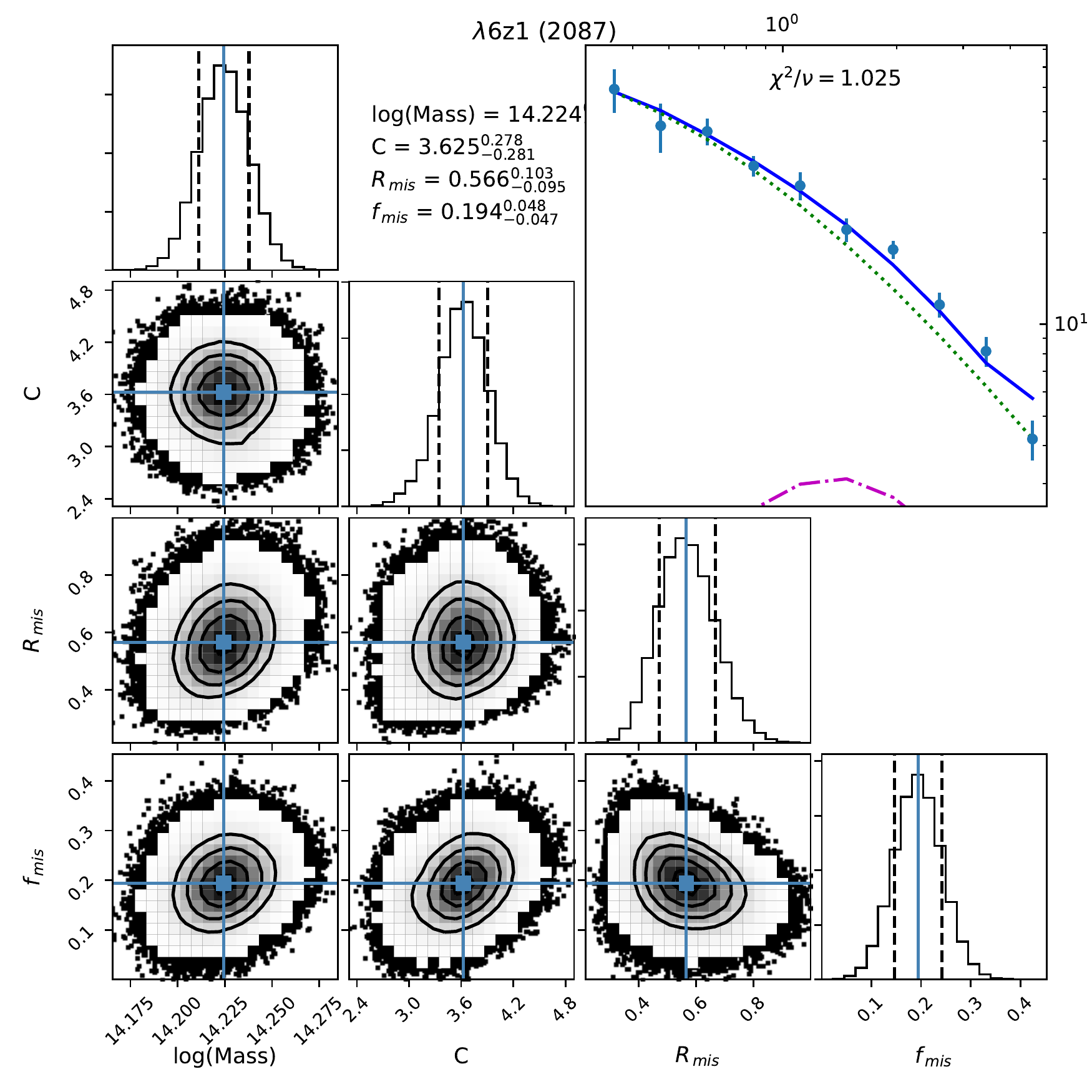}   \\
    Step 3 ~~~~~~~~~~~~~~~~~~~~~~~~~~~~~~~~~~~~~~~ Step 4 ~~~~~~~~~~~~~~~~~~~~~~~~~~~~~~~~~~~~~~~Step 5\\
    \includegraphics[scale=0.5]{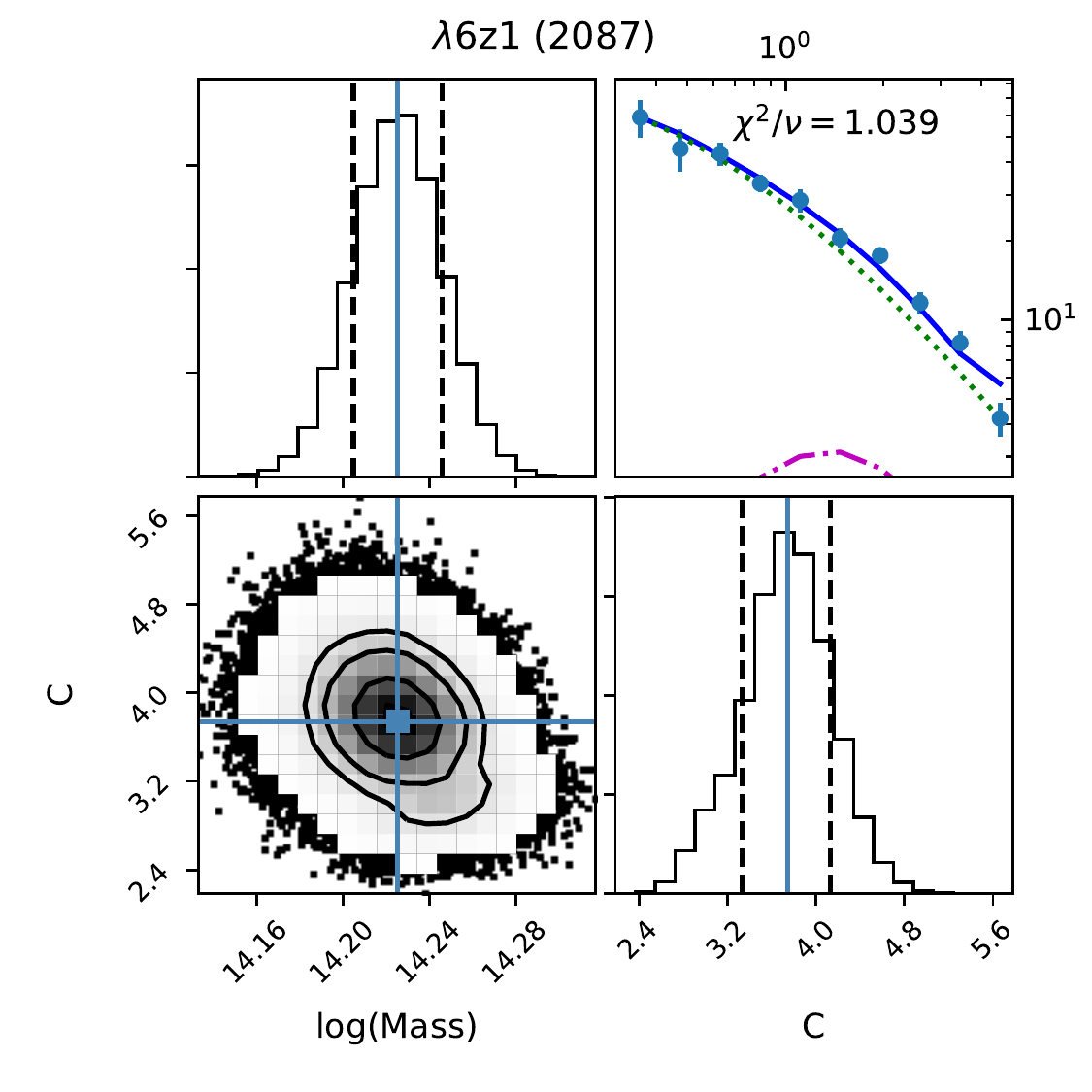}   
    \includegraphics[scale=0.5]{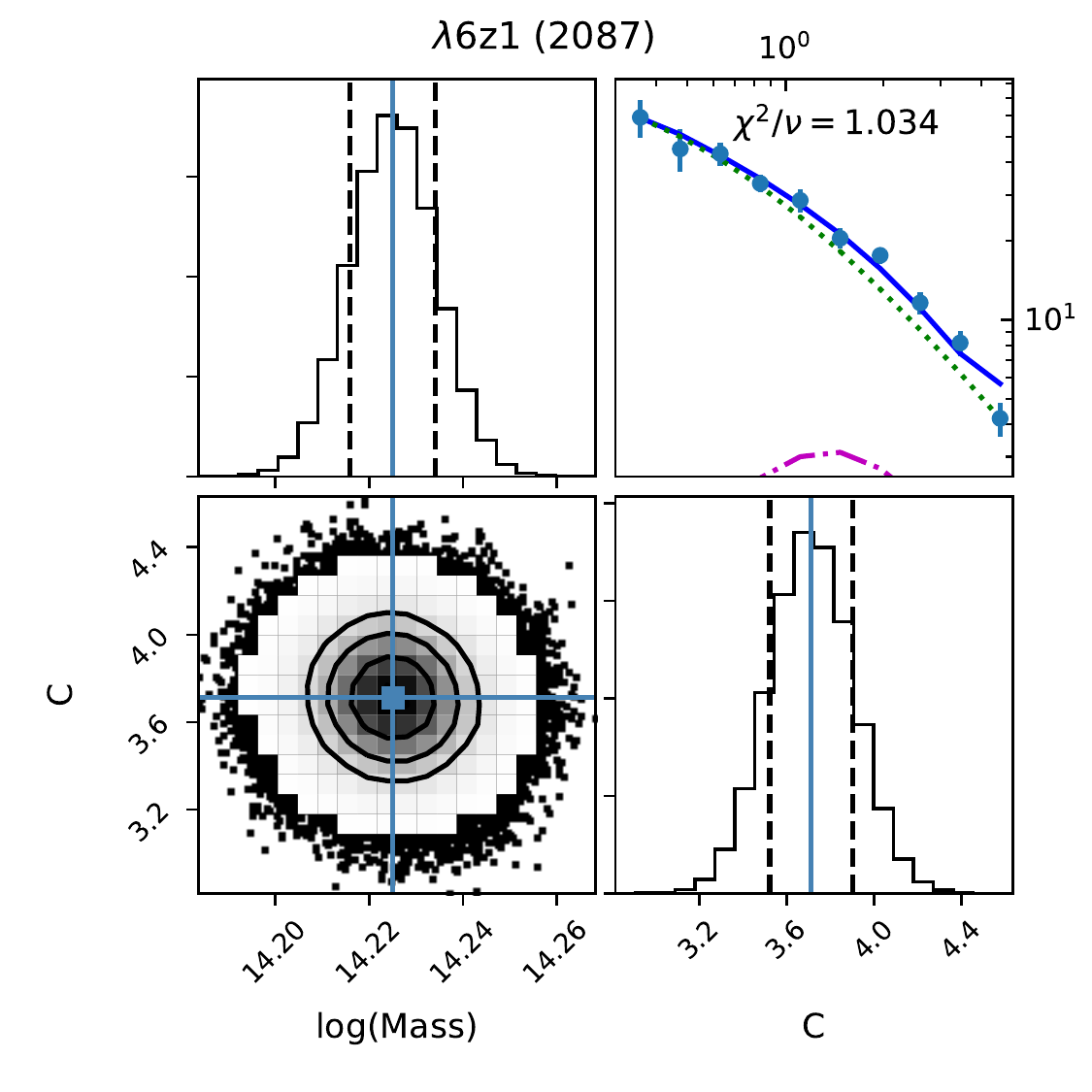}   
    \includegraphics[scale=0.8]{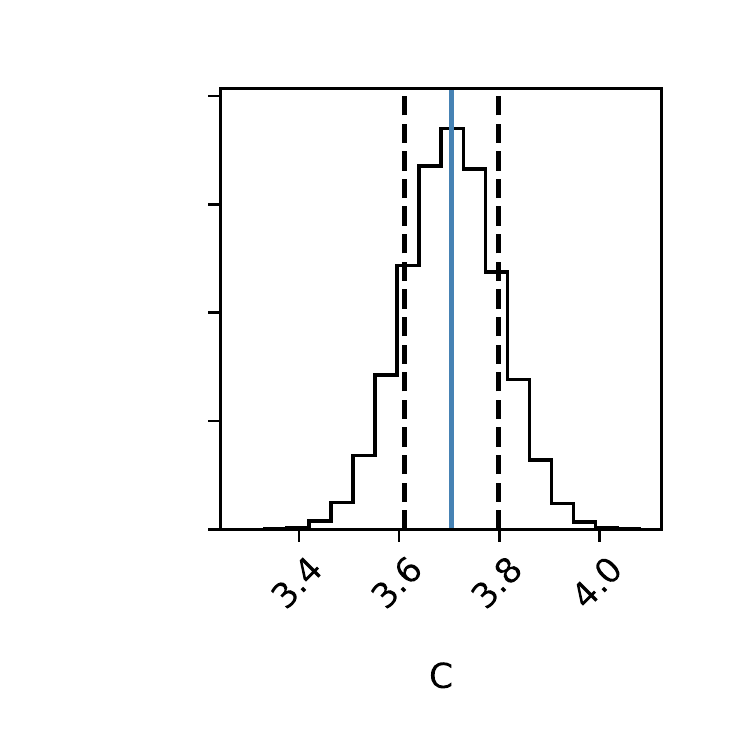}
    Step 5\\
    \includegraphics[scale=0.7]{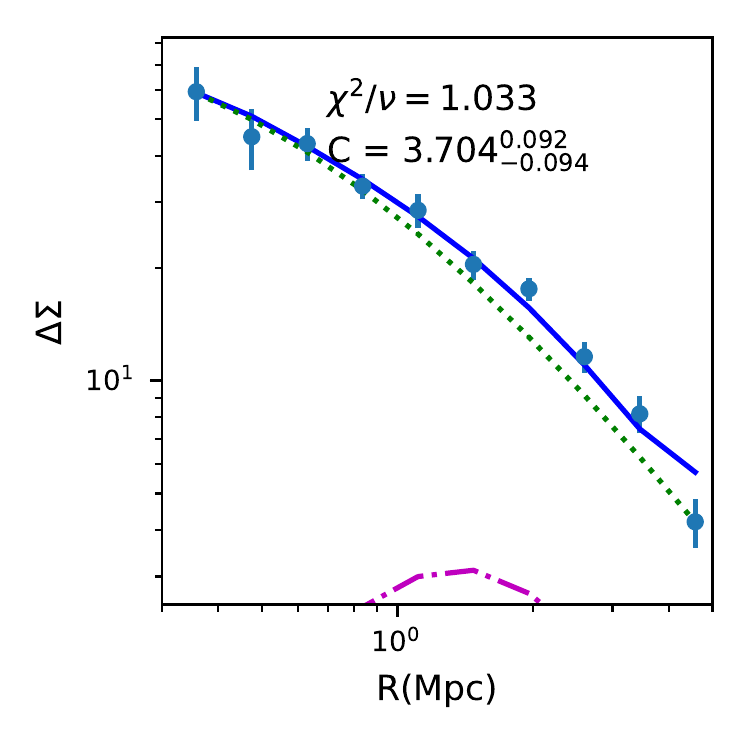}    
    \small
    \caption{The fitting steps of $\lambda 6z1$ as an example. The 1st to 5th panels corresponding to the step $1-5$ in the Sec.~\ref{sec:fit_steps}. The last panel shows the fitting result. In the top-right sub-panel of first two panels in first two rows, as well as the plot in the last row, the data and models are overlaid. The contribution from the central model, the nearby halo plus the mis-centering effect, and the total model are shown in green dotted curve, magenta dash-dotted curve and blue solid curve, respectively. In addition, the value of reduced $\chi^2$ is overlaid. In panels of the first row, best fitted parameters and errors are added to the top of the second
    column of sub-panels. The best fitted $c$ is also shown in the last panel. In each sub-panel with histogram, the median value and $1\sigma$ error are shown in solid cyan and dashed black vertical lines, respectively. }
    \label{fig:app}
\end{figure*}

In Fig.~\ref{fig:app}, we show the MCMC fitting result of the $z1\lambda6$ sample step by step as an example. 
The fitting result for step $1, 2, 3, 4, 5$ described in the last paragraph 
are shown in sequence as the panel $1-5$ in the figure, and the final fitting are shown in the last panel. 
The corresponding $\chi^2/\nu$ is labeled out for each fitting. 

\subsection{$c$-M relation fitting}
\label{sec:fit_pl_k16}

We fit the relation between the concentration and mass of halos with the model with upturn (called K16 model hereafter, shown in Equ.~\ref{equ:K16}, referring to Equ.~$24$ in \citealt{Klypin2016}), and the power-law model (called PL model hereafter, shown in Equ.~\ref{equ:pl}),

\begin{equation}
c({\rm{M}})=C_0~(\frac{\rm{M}}{10^{12}~\rm{M}_{\odot}/h})^{-\gamma}~[1+(\frac{\rm{M}}{M_0})^{0.4}],
\label{equ:K16}
\end{equation}

\begin{equation}
c({\rm{M}})=C_0~(\frac{\rm{M}}{M_0})^{-\gamma}.
\label{equ:pl}
\end{equation}

\begin{figure}[t]
\centering
\includegraphics[scale=0.6]{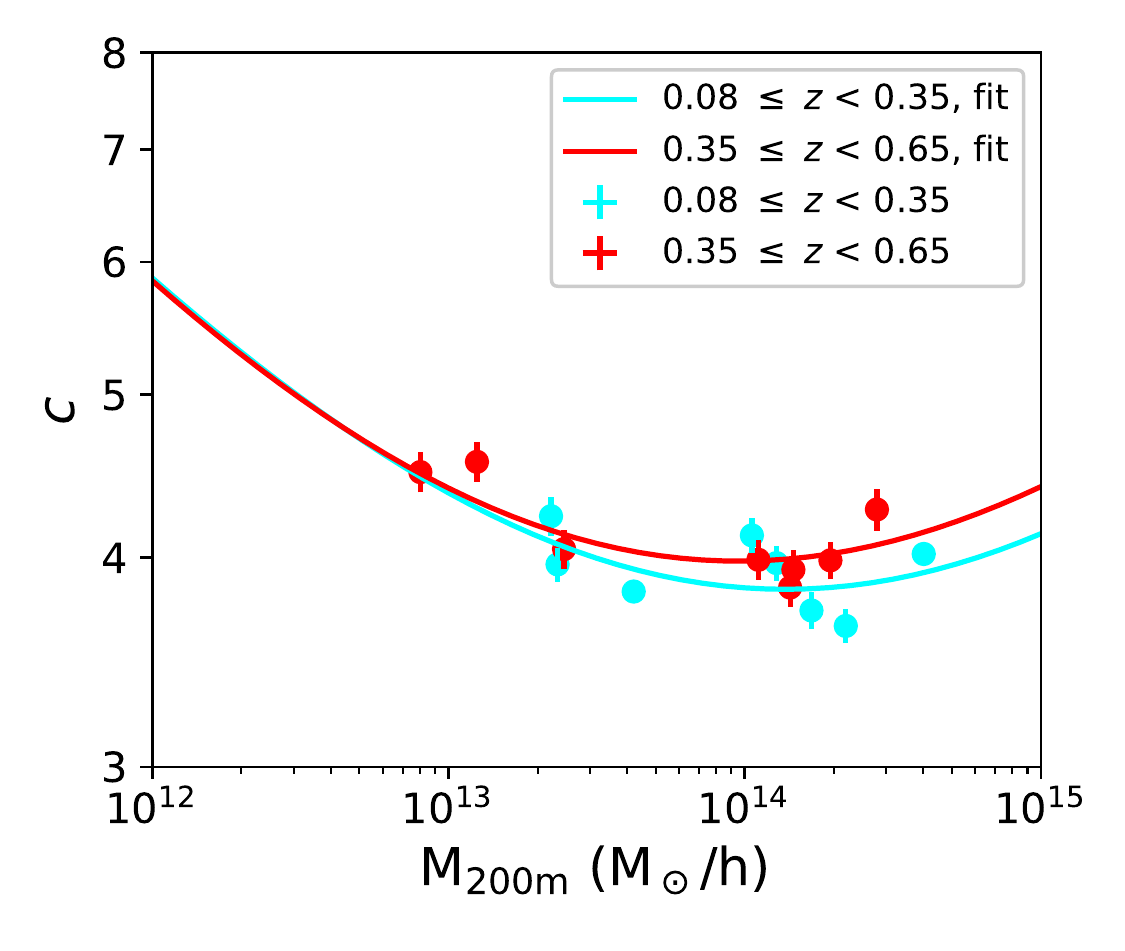}
\includegraphics[scale=0.6]{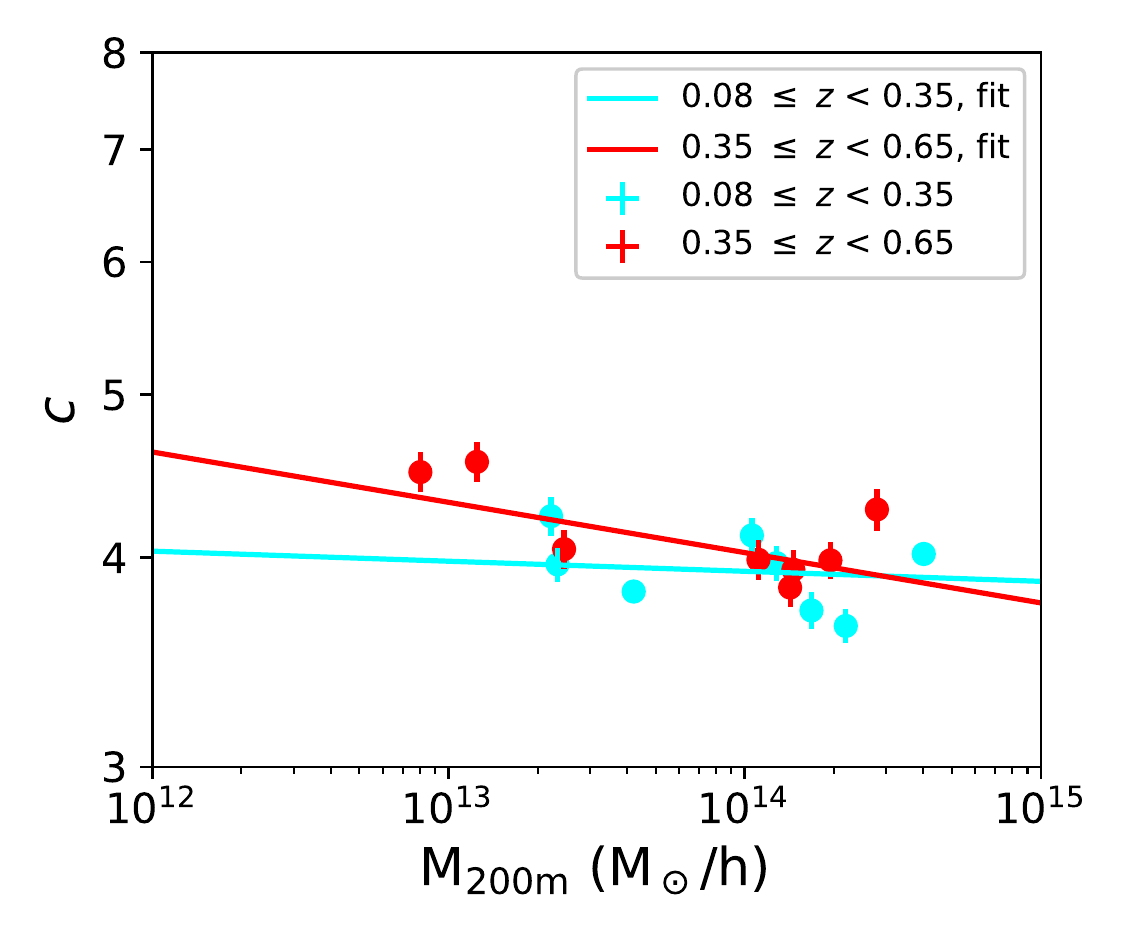}
\caption{The stacked signal and model of the weak lensing of clusters. The dots with error bars are
our results attained with DECaLS DR8 data, where the signal from the low redshift sample ($z1$ sample, $0.08\le z<0.35$) 
are shown in cyan, and the signal of high redshift sample ($z2$ sample, $0.35\le z<0.65$) in red. The 
solid curves are the best fitting model to the data,
which are shown in the same color with the corresponding data set. The model used in the upper panel is the K16 model, while in the lower panel the PL model. The best fitting parameters are listed in Tab.~\ref{tab:cmFit}. In $z1$ and $z2$ samples, the left three data points are from LOWZ and CMASS sample, respectively, while the other data are from redMaPPer sample.}
\label{fig:cmRelation}
\end{figure}

To remove the effect of redshift, we fit the data with two redshift bins separately. 
In this step, we firstly estimate parameters with the maximum likelihood method. 
Then, the best fitting parameters are obtained using the MCMC fitting method. 
We assume parameters ($C_0$, $\gamma$, $M_0$) as gaussian priors, 
whose central values are obtained from the 
maximum likelihood fitting, and FWHM as [$2, 0.13, 1$], respectively. 
In terms of the MCMC fitting setup, we take $50$~chains with the length of 
$30\,000$ steps and only last $20\,000$ steps are taken into account. 
In Fig.~\ref{fig:cmRelation}, we show the best fitting K16 model (top
panel) and the best fitting PL model (bottom panel) for samples in two 
redshift bins. The best fitting parameters are listed in Tab.~\ref{tab:cmFit}.
Both the reduced $\chi^2$ for $z1$ and $z2$ for K16 model are smaller than the
values obtained from the PL model. 
Thus, the K16 model is a better model to describe the $c$-M relation. 

\begin{table}[t]
    \centering
    \footnotesize
    \begin{tabular}{c|c|c|c|c|c}
\hline
Mod. & Samp. & $C_0$ & $\gamma$ & log$_{10}(M_0)$ & $\chi^2/\nu$ \\
\hline
K16 & $z1$ & $5.119_{-0.185}^{0.183}$ & $0.205_{-0.010}^{0.010}$ & $14.083_{-0.133}^{0.130}$ & $3.610$ \\
    & $z2$ & $4.875_{-0.208}^{0.209}$ & $0.221_{-0.010}^{0.010}$ & $13.750_{-0.141}^{0.142}$ & $1.243$ \\
\hline
PL  & $z1$ & $3.915_{-0.069}^{0.068}$ & $0.006_{-0.007}^{0.007}$ & $14.177_{-1.016}^{0.975}$ & $4.983$ \\
    & $z2$ & $4.053_{-0.267}^{0.266}$ & $0.030_{-0.009}^{0.009}$ & $13.905_{-0.989}^{0.996}$ & $2.880$\\
\hline
    \end{tabular}
    \caption{The best-fitting parameters and 1$\sigma$ error} of $c$-M relation. The $z1$ samples refer to the samples with $0.08\le z<0.35$, and $z2$ for $0.35\le z<0.65$. The reduced $\chi^2$ for the K16 model and PL model are listed in the last column.
    \label{tab:cmFit}
\end{table}

\section{Discussion}
\label{sec:discussion}
\subsection{Comparison with previous observations}
\label{subsec:compare_obs}

There are several works dedicated to measure the halo mass and concentration. In this section, we make a brief introduction of some previous works, followed by the comparison with our work. A summary of these works are shown in Tab.~\ref{tab:cm_obs}.

\begin{table*}[]
    \centering
    \begin{tabular}{c|c|c|c|c|c}
    \hline
    \hline
      $z$   & M$_{\rm halo}(10^{14}~\rm{M}_{\odot}/h)$ & N$_{\rm halo}$ & $\sigma_{\rm M}(10^{14}~\rm{M}_{\odot}/h)$  & $\sigma_c$ & Ref. \\
         \hline
      $0.1-0.3$   & $0.005-10^{\rm b}$  & $222\,699$    & $\sim 0.0867^{\rm b}$ & $\sim 2.997^{\rm b}$  & \citet{Mandelbaum2008} \\
      $0.1-0.6$   & $0.48-3.21^c$       & $1\,176$      & $0.185^c$             & $5.4^c$               & \citet{Covone2014} \\ 
      $0.19-0.69$ & $6.9-37.1^{\rm m}$  & $20$          &$0.095^c$              & $0.335^c$             & \citet{Umetsu2014} \\
      $0.19-0.89$ & $0.53-1.56^c$       & $19$          & $0.8^c$               &$1.0^c$                & \citet{Merten2015} \\
      $0.2-0.4$   & $0.05-2.00^c$       & $7\,425$      &$0.06^c$               & $0.91^c$              & low-$z$ in \citet{Shan2017} \\
      $0.4-0.6$   & $0.05-2.00^c$       & $27\,358$     &$0.05^c$               & $0.86^c$              & high-$z$ in \citet{Shan2017} \\
      $0.08-0.35$ & $0.05-5^{\rm m}$    & $301\,496$    &$0.0308^{\rm m}$       &$0.093^{\rm m}$        & low-$z$ in this work \\
     $ 0.35-0.65$ & $0.05-5^{\rm m}$    & $747\,376$    &$0.0394^{\rm m}$       &$0.108^{\rm m}$        & high-$z$ in this work \\
    \hline
    \hline
    \end{tabular}
    \caption{Compilation of previous works on measurement of $c$-M relation with observation. In the first three columns, the redshift range, the halo mass, and the halo number are shown in sequence. The uncertainties of the halo mass and concentration are listed in the fourth and fifth columns. The last column shows the corresponding reference. The superscript $^c$ ($^{\rm m}$, or $^{\rm b}$) denote the total mass enclosed within a sphere of radius r$_{200c}$ (r$_{\rm 200m}$, or r$_{\rm 200b}$), within which the mean density is $200$ times of the critical (mean matter background, or mean baryon background) density of the universe at the cluster redshift.}
    \label{tab:cm_obs}
\end{table*}

\citet{Mandelbaum2008} performed the statistical weak lensing analysis around the halos of $170\,640$ isolated galaxies, $38\,236$ groups and $13\,823$ maxBCG clusters. It includes the halo of galaxy size to cluster size, with the mass of $10^{12}-10^{15}~\rm{M}_{\odot}/h$. The estimated NFW concentration parameter $c_{200\rm b}$ decreases from $\sim 10$ to $4$ with halo mass. They fit the $c$-M relation with power-law function, and find the slope is in agreement with prediction of theory. However, the value of measured concentration is slightly smaller than the theoretical prediction and some other measurements. Millenium simulations predict the concentration becomes constant at higher mass range than the highest mass bin in their observation, and they only fit observation with power law relation in the work. Using WebPlotDigitizer\footnote{https://automeris.io/WebPlotDigitizer} \citep{Rohatgi2020}, we extract the data points in their Fig.~$5$. We discard the two leftmost and the rightmost data points because of the limited display. From these data points, we estimate the median uncertainties of the halo mass and concentration to be $\sim 8.7\times 10^{12}~\rm{M}_{\odot}/h$ and $\sim 3.00$, respectively.

\citet{Covone2014} obtain the stacked shear profile of $\sim 2\,000$ optical-selected 
galaxy clusters, wth the shear catalog from the CFHTLenS. The whole sample is divided into six 
richness bins to obtain the stacked shear profile separately. These bins corresponds to the 
M$_{200}$ from $0.48$ to $3.21\times10^{14}~\rm{M}_\odot/h$. The redshift coverage is 
$0.1-0.6$. According to the fitting result listed in the Tab.~$1$ therein, the median constraints 
of the halo mass and concentration is within $0.185\times10^{14}~\rm{M}_{\odot}$/h and $5.4$. 
In the model fitting of the measurements, they take into account the theoretical $\Lambda$CDM models 
of the hierarchical structure growth. The best fit slope is consistent with \citet{Duffy2008}, 
and the normalization differs within $1\sigma$ error.

In the work of \citet{Umetsu2014}, they made a joint shear-and-magnification weak-lensing 
analysis. A sample of $16$ X-ray-regular and $4$
high-magnification galaxy clusters selected from CLASH are taken into account. The redshift spans a range between $0.19$ and $0.69$. 
The result also agree with the $\Lambda$CDM prediction, especially the model of \citet{Meneghetti2013}. However, they didn't fit the measurements with the model with upturn. And the limited number of stacked clusters and the limited coverage of halo mass makes it difficult to distinguish the power-law model and the model with upturn.

With $19$ X-ray selected galaxy clusters from CLASH, \citet{Merten2015} derived a new constraint of the $c$-M relation. The redshift of this sample covers $0.19$ to $0.89$. 
The estimation of $c$-M relation agree with the theoretical estimation at $90\%$ confidence level. Using the combination of lensing reconstruction techniques of weak and strong lensing, they made a tight constraint of the mass and concentration, and the uncertainty is $0.8\times10^{14}~\rm{M}_{\odot}/h$ and $1.0$, respectively. The observation matches well with the full sample of \citet{Meneghetti2013}, and is similar with \citet{Bhattacharya2013}. However, there is also no fitting of the measurement with the model with upturn, and the limited sample impedes an accurate constraint of the $c$-M relation. 

In \citet{Shan2017}, they constrain the $c$-M relation with the redMaPPER clusters and LOWZ, CMASS galaxies. The halo mass is $5\times10^{12} – 2\times10^{14}~\rm{M}_{\odot}/h$. They only consider halos with the redshift of $0.2-0.4$ and $0.4-0.6$, including $7425$ and $27\,358$ halos respectively. With the large sample and accurate shear measurement from the CFHT Stripe 82 Survey, they constraint the uncertainty of halo mass within $0.06\times10^{14}~\rm{M}_{\odot}/h$, $0.05\times10^{14}~\rm{M}_{\odot}/h$, and the uncertainty of the concentration within $0.91$, $0.86$ for halos in two redshift bins. The measurements in two redshift bins are fitted with power-law model, and matches well with the simulation predictions (e.g., \citealt{Duffy2008}, \citealt{Klypin2016}). The comparison with power-law model and the model with upturn is not discussed in their work.

In this work, we focus on the stacked weak lensing signal of galaxy clusters in two redshift bins ( $0.08 \le z<0.35$ and $0.35\le z<0.65$), and only consider the halo with the mass in the range of $5\times10^{12} – 5\times10^{14}~\rm{M}_{\odot}/h$. The sample size is $301\,496$ and $747\,376$ for the low redshift and high redshift bin. Our uncertainties for halo mass is $3.08\times10^{12}~{\rm M}_{\odot}/h$ and $3.94\times10^{12}~{\rm M}_{\odot}/h$ in low and high redshift bins, respectively, while the uncertainty of concentration is $0.093$ and $0.108$.

Compared with previous observations, this work has the largest sample size, with more than 1 million halos taken into account. In addition, the least massive halo considered in our work ($\sim 5\times10^{12}~\rm{M}_{\odot}/h$) is lower than most of previous observation measurement, except the \citet{Mandelbaum2008} ($\sim 5\times10^{11}~\rm{M}_{\odot}/h$) and \citet{Shan2017} ($\sim 5\times10^{12}~\rm{M}_{\odot}/h$). However, we do not consider the very massive halos ($> 5 \times 10^{14}~\rm{M}_{\odot}/h$) as in \citet{Mandelbaum2008}, and \citet{Umetsu2014}. Furthermore, we have a quite wide redshift coverage ($0.08-0.65$), which covers the whole redshift area considered in the previous mentioned works, except for some high redshift halos in \citet{Umetsu2014} and \citet{Merten2015}. What is more import, we make a much tight constraint of the halo mass and the concentration, $2-10$ times better than previous works. 

\subsection{Comparison with previous cosmological simulations}
\label{subsec:compare_simu}
\begin{figure}[t]
    \centering
    \includegraphics[scale=0.6]{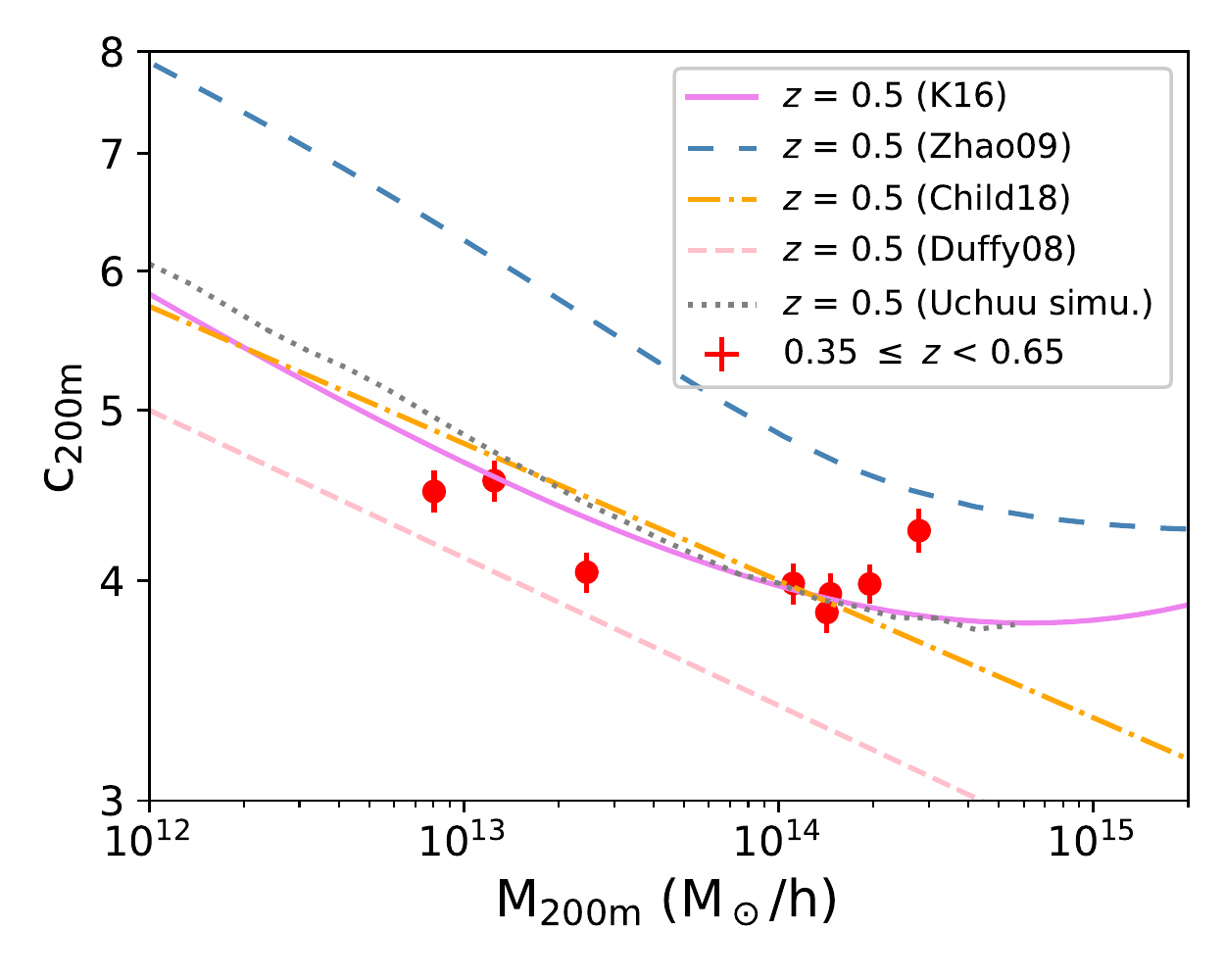}
    \caption{The comparison between our measurement for the high redshift sample ($z2$ sample, $0.35 \le z<0.65$ and cosmological predictions for halo at the redshift of $0.5$. The red data points refer our high-$z$ measurement,
    and lines for kinds of cosmological predictions of halos at the redshift of $0.5$. For the abbreviation, $K16$ refers to \citet{Klypin2016}, $Child18$ for \citet{Child2018}, $Duffy08$ for \citet{Duffy2008}, and $Uchuu~simu$ for \citet{Ishiyama2020}. The concentration of $Uchuu~simu$ used here is obtained with profile fitting. }
    \label{fig:cm_simu}
\end{figure}

In this section, we compare our measurement in the high redshift bin ($0.35\le z<0.65$) with simulations at the redshift of $0.5$, as shown in Fig.~\ref{fig:cm_simu}.
For the simulation in \citet{Zhao2009}, we obtain the $c$-M relation with the halo evolution web-calculator\footnote{http://www.shao.ac.cn/dhzhao/mandc.html}, with the power spectrum type set as $BBKS~1986~power ~spectrum$ \citep{Bardeen1986}.
We make the Kolmogorov-Smirnov test (K-S test) to estimate the performance of cosmological predictions. The estimation of $p$-value for the cosmological prediction suggested in \citet{Klypin2016}, \citet{Ishiyama2020} (called Uchuu simulation), \citet{Child2018}, \citet{Zhao2009}, \citet{Duffy2008} is $0.660$, $0.660$, $0.087$, $0.002$, $1.554\times 10^{-4}$, respectively. 
This comparison shows that the K16 model, as well as the Uchuu simulation fit our measurement better, compared with other mentioned cosmological models. 

\subsection{Correlation between parameters}

\begin{figure}[t]
    \centering
    \includegraphics[scale=0.75]{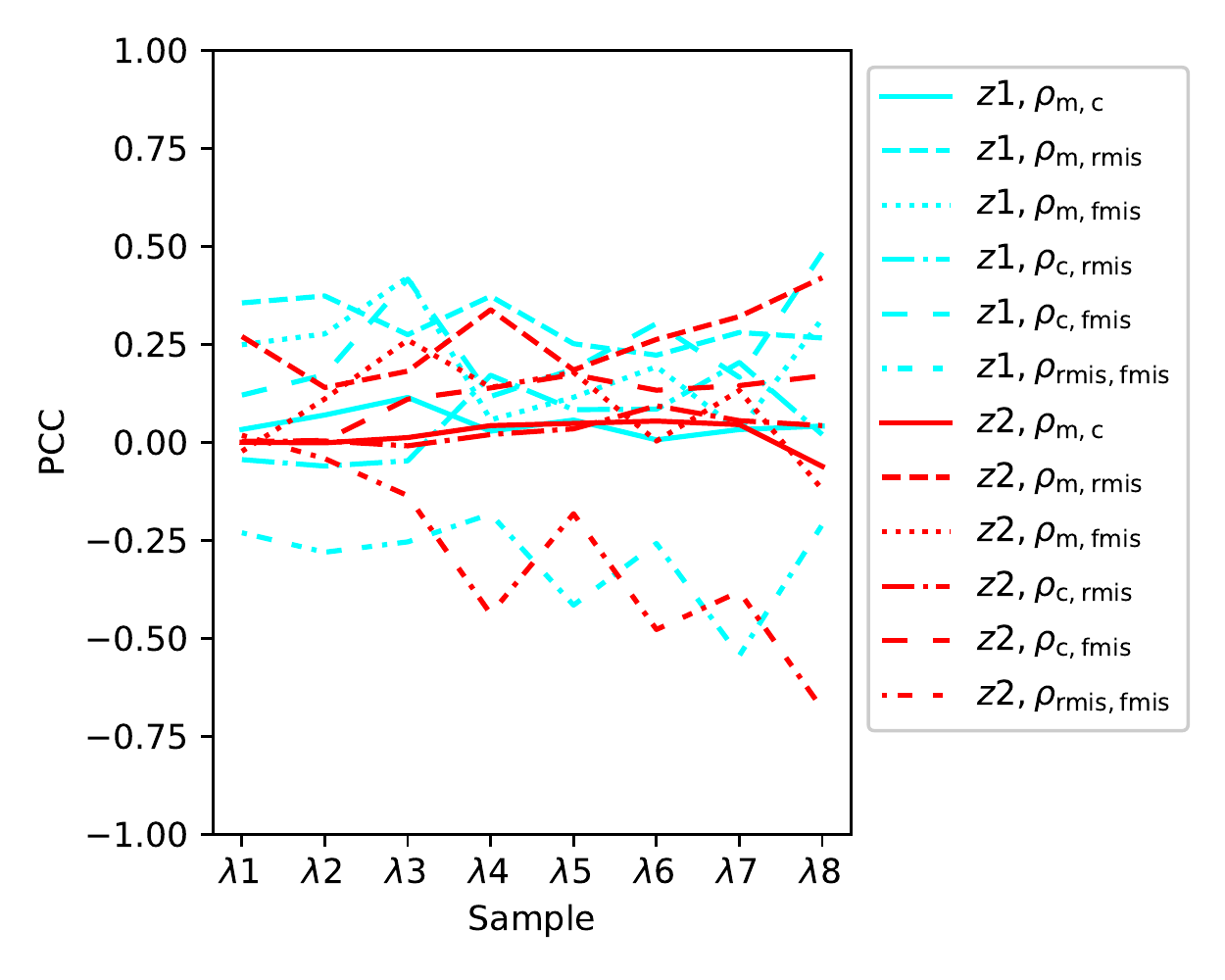}
    \caption{The PCC value for subsamples. }
    \label{fig:pcc}
\end{figure}

To quantify the correlation between the four free parameters, M$_{200}$, $c$, $r_{\rm mis}$, and $f_{\rm mis}$,
we calculate the Pearson correlation coefficient (PCC) for each subsample with Equ.~\ref{eq:pcc}. The MCMC output of step~$2$, as 
described in Sec.~\ref{sec:fit_steps}, are used. As shown in Fig.~\ref{fig:pcc}, the absolute PCC value is $<0.5$ for all sub-samples, except the one between the $r_{\rm mis}$ and $f_{\rm mis}$. The correlation between these two parameters is expected for both of them contribute the miscentering effect. 

\begin{equation}
\rho_{\rm X,Y}=\frac{\rm{Cov(X,Y)}}{\sigma_{\rm X} \sigma_{\rm Y}}.
\label{eq:pcc}
\end{equation}

\section{Conclusion}
\label{sec:conclusion}

With the DECaLS DR8 data, we extract and fit the stacked weak lensing signal of 
the halos in redMaPPer, LOWZ, and CMASS catalogs. 
Using the model of the central halo, nearby halo and miscentering effect, we model
the weak lensing signal and get the value of halo mass and concentration for
sub-samples in multiple redshift and mass bins. We obtain the $c$-M relation for the 
halos with mass range from $10^{13}$ to $10^{15}$~M$_{\odot}$ 
and redshift range from $z=0.08$ to $z=0.65$. 
Compared with power-law model, our fitting of the $c$-M relation prefers the K16 model \citep{Klypin2016}, which includes
a trend of upturn after the pivot point of $\sim 10^{14}~\rm{M}_{\odot}$. This is the first measurement of the $c$-M relation with DECaLS DR8 data. Our measurement shows the halo concentration with similar redshift decreases with the mass
increases, except for the upturn at massive end, which happens at lower mass for high redshift halo. For halos with similar mass, the halo at large redshift has larger concentration.

In this paper, we measure the $c$-M relation to detect the possible upturn at high-mass end.
Until now, there is still no consensus on the existence nor the reason of the upturn 
in the $c$-M relation. Some works claim it
comes from the unrelaxed dynamical state of dark matter halo \citep{Ludlow2012}.
Some others take the selection effect as the main contribution \citep{Meneghetti2013}.
Some more works explain the high concentration of massive clusters as a consequence
of the alignment of the major axis of the ellipsoidal halo and the line-of-sight 
(e.g., \citealt{Corless2009, Sereno2013, Limousin2013, Sereno2018}), with assuming 
the shape of dark matter halo as triaxial instead of spherical.

In addition, there is a possible degeneracy between $f_{\rm mis}$ and $c$. An
over-estimated factor of miscentered halo is likely to over-estimate the concentration. This would happen more frequently at high redshift. Thus, to 
get more accurate constraint of halo concentration, we need a better centering
strategy, such as the massive galaxies near the X-ray centroids \citep{George2012}. With the miscentering effect limited, the concentration would be measured more accurate, as well as the $c$-M relation. 

Furthermore, the upturn of the $c$-M relation can also be explained by the major merger of massive clusters, when the subhalo moves to the central part
radially and increase the concentration of the resulting halo \citep{Klypin2011,Prada2012,Klypin2016}.
In our measurement, the K16 model \citep{Klypin2016} with an upturn at massive regime shows a comparatively good fitting, suggesting that a trend of upturn exists. This measurement is important for the structure formation at high-mass end. We expect the next generation weak lensing surveys, such as Euclid \citep{Euclid2019}, LSST \citep{LSST2009}, CSST \citep{CSST2011}, will provide enough statistics to confirm and explain the existence of the upturn of the $c$-M relation at high-mass end.

\software{$Cluster~toolkit$ \citep{Smith2003,Eisenstein1998,Takahashi2012}, 
$CAMB$ \citep{Challinor2011,Lewis1999}, 
$SWOT$ \citep{Coupon2012}, 
$EMCEE$ \citep{emcee}, 
$WebPlotDigitizer$ \citep{Rohatgi2020}.}

\begin{acknowledgements}
We acknowledge support from the National Key R\&D Program of China (2016YFA0400703), the National Science Foundation of China (11721303, 11890693, 11973070) and the science research grants from the China Manned Space Project with NO. CMS-CSST-2021-A01. WX would like thanking Jin Wu, Siwei Zou, Junjie Jin, Yu
Qiu, Shu Wang, Zhiwei Pan for useful help and discussions during the development of this paper. HYS acknowledges the support from the Shanghai Committee of Science
and Technology grant No.19ZR1466600 and Key Research Program of Frontier Sciences, CAS, Grant No. ZDBS-LY-7013.

\end{acknowledgements}

\bibliographystyle{aasjournal}
\bibliography{main} 

\end{document}